\documentclass[a4paper,fleqn,usenatbib]{mnras}

\usepackage{mathptmx}

\usepackage[T1]{fontenc}
\usepackage{ae,aecompl}

\usepackage{graphicx}	
\usepackage{amsmath}	
\usepackage{amssymb}	

\title[Spirals in the quiescent accretion disc of V2051\,Oph]
      {Infrared photometry of the dwarf nova V2051 Ophiuchi: \\
        II - The quiescent accretion disc and its spiral arms
        \thanks{Based on observations
       obtained at the Southern Astrophysical Research (SOAR) telescope, which
       is a joint project of the Minist\'{e}rio da Ci\^{e}ncia, Tecnologia,
       Inova\c{c}\~{a}os e Comunica\c{c}\~{a}oes (MCTIC) do Brasil, the U.S.
       National Optical Astronomy Observatory (NOAO), the University of North
       Carolina at Chapel Hill (UNC), and Michigan State University (MSU).} }

\author[Baptista \& Wojcikiewicz]{
  Raymundo Baptista$^{1}$ \thanks{E-mail: raybap@gmail.com (RB)}
  and Eduardo Wojcikiewicz$^{1}$
\\
$^{1}$Departamento de F\'{i}sica, Universidade Federal de Santa Catarina,
 Campus Trindade, 88040-900 Florian\'{o}polis, SC, Brazil
}

\date{Accepted 2019 December 12. Received 2019 December 3;
  in original form 2019 October 7}

\pubyear{2019}

\begin{document}
\label{firstpage}
\pagerange{\pageref{firstpage}--\pageref{lastpage}}
\maketitle

\begin{abstract}
  
 We report the analysis of time-series of infrared $JHK_s$ photometry of
 the dwarf nova V2051\,Oph in quiescence with eclipse mapping techniques
 to investigate structures and the spectrum of its accretion disc.
 The light curves after removal of the ellipsoidal variations caused by
 the mass-donor star show a double-wave modulation signalling the
 presence of two asymmetric light sources in the accretion disc.
 Eclipse maps reveal two spiral arms on top of the disc emission, one at
 $R_1= 0.28\pm 0.02 \,R_\mathrm{L1}$ and the other at $R_2= 0.42\pm 0.02
 \,R_\mathrm{L1}$ (where $R_\mathrm{L1}$ is the distance from disc centre to
 the inner Lagrangian point), which are seen face-on at binary phases
 consistent with the maxima of the double-wave modulation. The wide open
 angle inferred for the spiral arms ($\theta_s= 21^o \pm 4^o$) suggests
 the quiescent accretion disc of V2051\,Oph has high viscosity. The
 accretion disc is hot and optically thin in its inner regions
 ($T_\mathrm{gas}\sim 10-12 \times 10^3\,K$ and surface densities $\sim
 10^{-3}-10^{-2}\,g\,cm^{-2}$), and becomes cool and opaque in its outer
 regions.

\end{abstract}

\begin{keywords}
  accretion, accretion discs -- binaries: eclipsing -- binaries:close --
  stars: dwarf novae -- stars: individual (V2051\,Ophiuchi)
\end{keywords}



\section{Introduction}

Dwarf novae are short period interacting binaries where a late type star
overfills its Roche lobe and transfers matter to a companion white dwarf
(WD) via an accretion disc. They show days-long outbursts where the disc
brightens by 2-5 mag and which reccur on timescales of days to weeks.
Aside of the normal outbursts, the sub-class of the SU\,UMa stars show
stronger, longer and more regular superoutbursts characterized by the
appearance of superhumps, believed to be the result of the tidal
interaction of the donor star with a slowly precessing, elliptical outer
ring which develops when the disc expands beyond the 3:1 resonance
radius \citep[e.g.,][]{whitehurst88,ho90,lubow94}. Accretion discs in
dwarf novae are also prone to the appearance of tidally-induced, wide
open spiral density waves if the disc is hot and/or highly viscous
\citep[e.g.,][]{Sawadaetal86,Steeghsetal97,Stehle99}.

Dwarf nova outbursts are thought to be driven either by bursts of
increased mass transfer rate \citep[Mass Transfer Instability Model,
MTIM, e.g.,][]{Bathpringle1981} or by a thermal-viscous disc instability
\citep[Disc Instability Model, DIM, e.g.][]{Lasota2001}. MTIM interprets
the outburst as the response of a constant, high viscosity
 \footnote{here we adopt the prescription of \cite{ss} for the accretion
   disc viscosity, $\nu = \alpha c_s H$, where $\alpha$ is the
   non-dimensional viscosity parameter, $c_s$ is the local sound speed
   and $H$ is the disc scaleheight.}
disc \citep[$\alpha\sim 0.1-1$, from the decline timescale of
outbursting dwarf novae, e.g.,][]{w95} to a burst of mass transfer from
the donor star. On the other hand, DIM predicts matter accumulates in a
cool, low viscosity disc during quiescence ($T_\mathrm{ef} < T_\mathrm{crit1}
\sim 7000\,K \,\,, \,\,\alpha_{\rm cool}\sim 10^{-2}$) which switches to a
hot, high-viscosity regime during outbursts ($T_\mathrm{ef} > T_\mathrm{crit2}
\sim 10000\,K\,\,, \,\,\alpha_\mathrm{hot}\simeq10\, \alpha_\mathrm{cool}
\sim 0.1$). This limit-cycle scheme implies that the disc temperatures
must be $T_\mathrm{ef}<T_\mathrm{crit1}$ in quiescence and $T_\mathrm{ef}>
T_\mathrm{crit2}$ during outbursts, making the comparison of observed
quiescent and outburst disc temperatures with $T_\mathrm{crit1}$ and
$T_\mathrm{crit2}$ a key test for DIM.

V2051\,Oph is a short-period ($P_\mathrm{orb}=90$ min) eclipsing
dwarf nova showing large-amplitude flickering (random brightness
fluctuations of $0.1-1$ mag), deep eclipses ($\Delta M_\mathrm{B}\simeq
2.5$\,mag) and a plethora of different eclipse profiles
\citep{Warner83,Cook83,Warner87}.  Superoutbursts were observed and
superhumps were detected by \citet{Kiyota98}, \citet{Vrielmann03} and
\citet{Pattersonetal03}, implying that V2051 Oph is an SU UMa type dwarf
nova. \citet{Baptistaetal98} used HST and ground-based observations to
constrain the binary parameters, finding a mass ratio of
$q = 0.19\pm 0.03$, an inclination of $i=83.3^\circ\pm1.4^\circ$,
and precise values for the mass and radius of both stars.

The multi-color photometric study of \citet{Vrielmann02} and the
UV-optical spectroscopic study of \citet{Saito06} both suggest that
the quiescent accretion disc of V2051\,Oph has a hot, optically thin
chromosphere responsible for the emission lines, possibly enveloping
a cooler and opaque underlying disc. \cite{Papadaki08} performed
Doppler tomography of V2051\,Oph in quiescence, two days after the end
of a superoutburst in 1999. Their Balmer and He\,I tomograms show two
asymmetric arcs at opposite sides of the accretion disc, reminiscent
of the spiral structures seen in Doppler tomograms of IP\,Peg in
outburst \citep[e.g.,][]{Steeghsetal97}, which they instead associated
with the bright spot and with the superhump light source.
H\,I, He\,I and O\,I Doppler tomography of V2051\,Oph in quiescence
several months after its previous outburst also shows two asymmetric
arcs extended in azimuth at opposite sides of the accretion disc,
which were then interpreted as being tidally-induced spiral arms
\citep{Rutkowski}.

An eclipse mapping investigation of the flickering sources by \citet{bb04}
reveal that the low-frequency flickering arises mainly in the
overflowing/penetrating gas stream, while the high-frequency flickering
originates in the accretion disc, suggesting a quiescent high viscosity
parameter of $\alpha \simeq 0.1-0.2$ at all disc radii if the disc
flickering is caused by magnetohydrodynamic turbulence \citep{ga92}.
\citet{Baptistaetal07} investigated the changes in disc structure
along two outbursts of V2051\,Oph and found that (i) the
disc shrinks at outburst onset, with enhanced emission along the
stream trajectory close to the circularization radius, (ii) the
cooling wave characterizing the outburst decline accelerates as it
travels towards disc centre, and (iii) for distances lower than
120\,pc the inferred disc brightness temperatures suggest that the
disc is cooler than $T_\mathrm{crit2}$ everywhere, therefore excluding
DIM as a viable explanation for its outbursts. The combined results
of \citet{bb04} and \citet{Baptistaetal07} are in clear contradiction
with DIM predictions, making V2051\,Oph an important challenge
to the prevailing DIM ortodoxy.

\citet[][hereafter Paper\,I]{PaperI} modelled the ellipsoidal variations
caused by the distorted mass-donor star of V2051\,Oph to infer its
$JHK_s$ fluxes, deriving a spectral type of $M(8.0\pm 1.5)$, an
equivalent blackbody temperature of $T_\mathrm{BB}=(2700\pm270)\,K$,
and a photometric parallax distance of $d_\mathrm{MS}=(102\pm16)$ pc
to the binary.  Here we apply eclipse mapping techniques to the infrared
$JHK_s$ light curves of Paper\,I, after subtraction of the ellipsoidal
variations caused by the mass donor star, in order to investigate the
structures and the broad-band infrared spectra of the quiescent
accretion disc of V2051\,Oph. This paper is organized as follows.
Sect.\,2 briefly describes the observations. Sect.\,3 reports the data
analysis and the results, which are discussed in Sect. 4 and summarized
in Sect. 5.

\section{The Observations and The Data}

Time-series of $JHK_s$ photometry of V2051\,Oph were collected with the
OSIRIS infrared imager at the 4.1\,m SOAR Telescope along the night of
2013 June 20 while the star was in quiescence. The observations framed two
consecutive orbital cycles in the $K_s$ band followed by about 1.5 orbital
cycles in the $H$ and $J$ bands each. The reader is referred to Paper\,I
for the details of the observations, data reduction and flux calibration
procedures. Although the $JHK_s$ observations are not simultaneous, they
were obtained over a short time interval in which the object remained at
the same brightness level.  Therefore, we are confident that the combined
$JHK_s$ surface brightness distributions (see Sect.\ref{eclipsemap}) do
represent the broad-band infrared spectra of the V2051\,Oph accretion disc
at that epoch.

In Paper\,I we modelled the ellipsoidal variation caused by the distorted
mass-donor star. For the remainder of this paper, the light curves under
analysis are those after subtraction of the mass-donor star contribution.
These light curves were phase-folded according to the linear plus sinusoidal
ephemeris of \citet{Baptistaetal2003},
\begin{eqnarray}
  T_\mathrm{mid}= \mathrm{BJDD}\,\, 2\,443\,245.977\,45 + 0.062\,427\,8629
  \times E \nonumber \\
  \mbox{} + 20 \times 10^{-5}\,\cos\left[ 2\pi
  \left(\frac{E - 120\times 10^3}{127\times 10^3}\right) \right]\, d \, ,
  \label{eq:ephem}
\end{eqnarray}
where $T_\mathrm{mid}$ is the white dwarf mid-eclipse time and $E$ is the
binary cycle. With this ephemeris the eclipses are centred at phase zero.

\section{Data analysis and results}
\label{analysis}

\subsection{The GAIA distance estimate}
\label{distance}

The GAIA Data Release 2 \citep{gaia16,gaia18,lindegren} yields a
trigonometric parallax of $\pi= 8.90\pm 0.07$\,mas for V2051\,Oph, which
translates into a distance of $(112.3\pm 0.9)$\,pc, consistent with
the photometric parallax distance estimates of Paper\,I at their
1-$\sigma$ limit.

\citet{Baptistaetal07} found that, in order to reconcile the observed
disc brightness temperatures at the 2002 August outburst maximum with the
critical temperatures required by DIM, V2051\,Oph should be at a distance
of at least 120\,pc. The above trigonometric parallax distance is lower
than this limit at a confidence level above the 8-$\sigma$ limit.
In fact, at a distance of 112\,pc even the stronger 2000 July outburst
occurred at disc temperatures below those required by DIM everywhere and
along the whole outburst \citep{Baptistaetal07}. Therefore, unless the
accretion disc of V2051\,Oph is optically thin during outburst -- in
which case the inferred B-band blackbody brightness temperatures of
\citet{Baptistaetal07} underestimate the true disc effective temperatures
(e.g., see Sect.~\ref{radtemp}) -- the GAIA trigonometric parallax
distance estimate excludes DIM as a viable mechanism to explain the
outbursts of V2051\,Oph.

The good agreement between the trigonometric and the photometric parallax
distance estimates indicates that the mass-donor star in V2051\,Oph is
indistinguishable from an isolated, main sequence star of similar mass,
consistent with the conclusions of \citet{Kniggeetal2011}.

\subsection{Analysis of the light curves}
\label{curves}

\begin{figure}
\centering
  \includegraphics[width=0.37\textwidth,angle=-90]{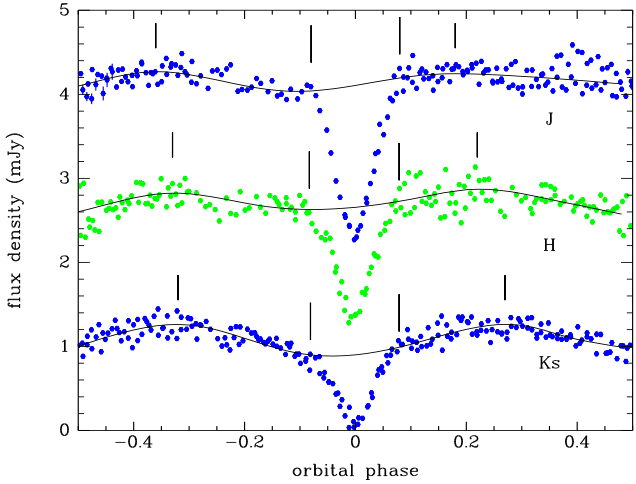}
  \caption{Phase-folded $JHK_s$ light curves of V2051 Oph (dots) and the
  best-fit double-wave spline function to the out-of-eclipse regions (solid
  line). The $J$ and $H$ light curves are vertically displaced, respectively,
  by 1 and 2 mJy for visualization purposes. Small vertical ticks show the
  phases of maximum of the double-wave modulation; large vertical ticks mark
  the phases of eclipse ingress/egress.}
   \label{fig:curvas}
\end{figure}

Figure\,\ref{fig:curvas} shows the $JHK_s$ orbital light curves of V2051\,Oph
after subtraction of the ellipsoidal variation caused by its mass donor star.
The light curves show a double-wave modulation (at $P_\mathrm{orb}/2$), the
amplitude of which increases with increasing wavelength. The maxima of the
modulation also move towards later phases with increasing wavelength.
This modulation is asymmetric in the sense that the maxima at negative
phases are stronger than those at positive phases, particularly in the $J$
and $K_s$ bands. Thus, we may discard an explanation in terms of possible
underestimation (overestimation) of the ellipsoidal variation of the
mass-donor star because this would lead to a residual double-wave modulation
of equal maxima (minima) centred at phases $\pm 0.25$, differently from
what is observed. The best-fit spline function to the out-of-eclipse
fluxes are shown as solid lines in Fig.\,\ref{fig:curvas}.
Measured peak-to-peak amplitudes, $\Delta f$, and phases of maxima from
these spline fits are collected in Table\,\ref{tab:parameters}.
%
\begin{table}
  \centering
  \caption{Parameters derived from the $JHK_s$ light curves}
  \label{tab:parameters}
  \begin{tabular}{lccc} 
   \hline
   Parameter & $J$ & $H$ & $K_s$ \\
   \hline
   $\phi_\mathrm{max1}$ &  $-0.36(2)$ &  $-0.33(3)$ &  $-0.32(2)$ \\
   $\phi_\mathrm{max2}$ &  $+0.18(3)$ &  $+0.22(3)$ &  $+0.27(1)$ \\
   $\Delta f$ (mJy)   &  0.23(3)    &  0.29(1)    &  0.37(1)  \\
   $1/2\,\Delta\phi_E$ &  0.080(2) & 0.081(2) & 0.082(3) \\
   $R_d/a$            &  0.26(2)  & 0.27(2)  & 0.28(2)  \\
   $R_d/R_\mathrm{L1}$  &  0.40(2) &  0.41(2) &  0.42(3)  \\
   \hline
  \end{tabular}
\end{table}

We also used the spline fits to measure the half-width of the eclipse
at each passband, $\Delta\phi_\mathrm{E}$, finding the phases where the
fitted spline deviates from the data by 1$\sigma$ of the local,
out-of-eclipse median flux, and dividing the resulting phase range by
two. We then applied the method of \cite{sulkanen} to estimate the disc
radius at each passband from the corresponding $\Delta\phi_\mathrm{E}$
value. By assuming a spherical secondary star it is possible to
derive the disc radius $R_d$ in units of the orbital separation 
$a$ from the analytical expression,
\begin{equation}
\frac{R_d}{a} = 
\sin(2\pi\Delta\phi_E)\sin i - \sqrt{(R_2/a)^2 - \cos^2 i} \,,
\end{equation}
where $R_2/a$ is the radius of a sphere containing the same 
volume as the Roche lobe of the secondary star, given by the 
relation \citep{eggleton},
\begin{equation}
\frac{R_2}{a}= \frac{0.49\,q^{2/3}}{0.6\,q^{2/3}+\log(1+q^{1/3})}\, .
\end{equation}
where $q=M_2/M_1$ is the binary mass ratio. Values of $R_d/a$ are
transformed to $R_d/R_\mathrm{L1}$ assuming $R_\mathrm{L1}/a= 0.66\pm 0.01$
\citep{Baptistaetal98}. The resulting values are listed in
Table\,\ref{tab:parameters} (with uncertainties given in parenthesis).
Since the differences are within the uncertainties, there is no
evidence that the disc radius varies with wavelength.

Previous estimates of V2051\,Oph disc radius from the position of the
bright spot at disc rim lead to $R_\mathrm{bs}/R_\mathrm{L1}= 0.56 \pm 0.02$
in a low brightness state \citep{Baptistaetal98} and
$R_\mathrm{bs}/R_\mathrm{L1}= 0.48 \pm 0.02$ in quiescence
\citep{bb04,Baptistaetal07}. It is not clear if the inferred smaller disc
radius reflects real changes in disc structure with time (i.e., changes
in mass transfer rate) or if it is a consequence of systematic differences
in measuring technique.

\subsection{Eclipse mapping}
\label{eclipsemap}

The eclipse mapping method is an indirect imaging technique which translates
the information in the eclipse shape into a map of the surface brightness
distribution of the occulted disc regions \citep{Horne85}. Because the
one-dimensional data light curve do not fully constrain a two-dimensional
surface brightness map, a maximum entropy procedure \citep[e.g.,][]
{Skilling87} is used to select, among all possible solutions, the one that
maximizes the entropy of the eclipse map with respect to a smooth default
map \citep{Baptista01,Baptista16}. The usual choice of default map leads
to the most nearly axi-symmetric map that fits the data.

In order to improve the signal-to-noise ratio and to minimize the
influence of flickering, the light curve of each passband was binned to
phase resolutions of $\delta\phi= 0.007$ and 0.025, respectively during
eclipse ($\Delta\phi= -0.1,+0.1$) and outside of eclipse. The median flux
was computed for each bin; the median of the absolute deviations with
respect to the median was taken as the corresponding uncertainty at each
bin.  Maximum entropy eclipse mapping techniques were applied to the
$JHK_s$ binned light curves to solve for a map of the surface brightness
distribution of the accretion disc and for the flux of an additional
uneclipsed component in each passband.

At the high inclination of V2051\,Oph,
flaring of its accretion disc (at half-opening angle $\beta$) may lead
to perceptible differences in apparent brightness between the near and
far sides of the disc (seen at effective angles of $i+\beta$ and
$i-\beta$, respectively), and ignoring this effect may have a
significant impact in the interpretation of eclipse mapping results.
Therefore, as a first step, we used a 3D version of the eclipse mapping
method in which the eclipse map is a conical surface with side
$2\,R_\mathrm{L1}$ centred at the WD position and inclined at a
half-opening angle $\beta$ with respect to the orbital plane, plus a
circular rim orthogonal to the orbital plane at a distance $R_\mathrm{d}$
($< R_\mathrm{L1}$) from the disc centre.  The disc rim radius
$R_\mathrm{d}$ and half-opening angle $\beta$ are additional free
parameters of the method, which can be inferred from an entropy
landscape procedure \citep[][and references therein]{Baptistaetal16}:
eclipse maps are computed for a range of $(R_d,\beta)$ pair os values
and the resulting space of parameters is searched for the combination
that yields the eclipse map of highest entropy. The resulting space of
parameters is well behaved with a single, well defined entropy maximum
in each case. We find $R_\mathrm{d}/R_\mathrm{L1}= 0.42 \pm 0.05$ independent
of passband, in good agreement with the disc radius estimated from the
width of the eclipse. We also find $\beta= 0.0\degr \pm 0.5\degr$,
indicating that the accretion disc of V2051\,Oph is geometrically thin.

\begin{figure*}
  \centering
  \includegraphics[width=0.8\textwidth,angle=-90]{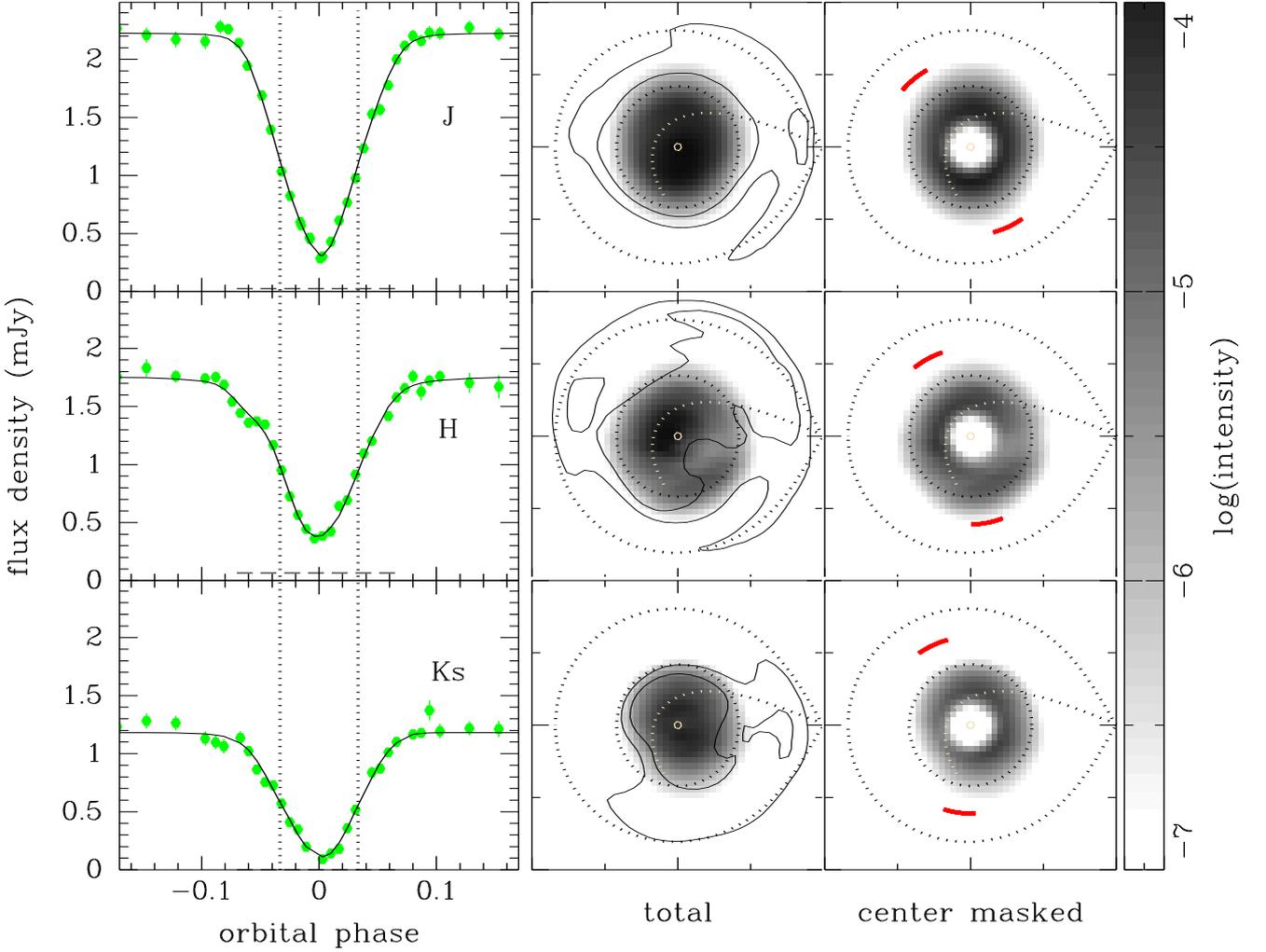}
  \caption{Left-hand: $JHK_s$ data (dots with error bars) and model (solid
   lines) light curves. Horizontal dashed lines indicate the uneclipsed flux
   in each passband. Vertical dotted lines mark the ingress/egress phases of
   the WD eclipse.  Middle: eclipse maps in a logarithmic grayscale. Regions
   inside the two solid contour lines are above the 3$\sigma$ and 5$\sigma$
   confidence levels, respectively. A circle marks the position/size of the WD
   at disc center. Dotted lines depict the primary Roche lobe, the gas stream
   trajectory and a circle of radius $0.42\,R_\mathrm{L1}$.  Right-hand: the
   eclipse maps with their central regions ($R<0.2\,R_\mathrm{L1}$) masked for
   a clearer view of the outer disc regions. Small arcs (red) indicate the
   azimuths of maximum emission of the double wave modulation in each case.
   The rightmost vertical bar shows the logarithmic intensity grayscale;
   brighter regions are darker.}
    \label{fig:maps}
\end{figure*}

Given that the disc half-opening angle is negligible, we decided to
proceed with a standard, 2D version of the eclipse mapping code in which 
the eclipse map is a flat cartesian grid of $51\times 51$ pixels centred at
the WD position with side $2 R_\mathrm{L1}$. The eclipse geometry is
defined by the mass ratio $q$ and the binary inclination $i$, and the scale
of the map is defined by the value of $R_\mathrm{L1}$. We adopted $q=0.19\pm
0.03$, $i=83.3\degr \pm 1.4\degr$ and $R_\mathrm{L1}= (0.422\pm 0.015)\,
R_\odot$, which correspond to a WD eclipse width of $\Delta\phi= 0.0662\pm
0.0002$ \citep{Baptistaetal98}. This combination of parameters ensures that
the WD is at the centre of the map.
Because this version of the eclipse mapping method does not take into
account out-of-eclipse brightness changes, these were removed by dividing
each light curve by the corresponding best-fit spline function
(Sect.\ref{curves}), and scaling the result to the average out-of-eclipse
flux level. The resulting light curves are shown in the left-hand panels
of Fig.\ref{fig:maps}.

Our eclipse mapping code implements a scheme of double default functions,
$D_+\,D_-$, simultaneously steering the solution towards the most nearly
axi-symmetric map consistent with the data ($D_+$), and away from the
criss-crossed arcs along the edges of the shadow of the occulting, mass
donor star ($D_-$) \citep{Spruit1994,Baptista2005,Baptista16}. It is
optimized to recover asymmetric structures in eclipse maps such as spiral
arms and enhanced gas stream emission. The positive default function is a
polar Gaussian with radial and azimuthal blur widths of
$\Delta r= 0.02\,R_\mathrm{L1}$ and $\Delta \theta = 30^\circ$, respectively.
The negative default function is a Gaussian along the ingress/egress arcs
of phase width $\Delta \phi = 0.01$.

The uncertainties in the eclipse maps were derived from Monte Carlo
simulations with the light curves, generating a set of 50 randomized eclipse
maps \citep[see][]{Ruttenetal1992}.  These are combined to produce a map
of the standard deviations with respect to the true map. A map of the
statistical significance (or the inverse of the relative error) is obtained
by dividing the true eclipse map by the map of the standard deviations
\citep{Baptista2005}. The uncertainties obtained with this procedure are
used to plot the confidence level contours in the eclipse maps of
Fig.~\ref{fig:maps} and to estimate the uncertainties of the
spatially-resolved disc spectra (Figs.~\ref{fig:spectra} and
\ref{fig:arms}).
%
%
The uncertainties in the eclipse geometry (derived from the errors in
the binary parameters) are negligible in comparison with the statistical
uncertainties affecting the eclipse maps. An additional systematic
uncertainty of 10 per cent was added in quadrature to the uncertainty of
each data point in the spatially-resolved disc spectra to account for the
errors introduced by the subtraction of the contribution of the mass donor
star from the light curves. This has a minor effect in the results as the
statistical uncertainties of the spatially-resolved spectra are dominant,
in most of the cases being around the 15-20 per cent level.

\subsection{Spiral structures in the accretion disc}
\label{spirals}

Data and model light curves and corresponding eclipse maps in a logarithmic
grayscale are shown in Fig.~\ref{fig:maps}.
The uneclipsed component corresponds to about 1, 4 and less than 1 per cent
of the total flux, respectively in the $J$, $H$ and $K_s$ bands, indicating
no significant contribution from a stellar disc wind to the infrared light
of V2051\,Oph in quiescence.
The statistical significance of the structures in the eclipse maps are at
the 3$\sigma$ and above the 5$\sigma$ confidence levels, respectively for
the outer (fainter) and inner (brighter) disc regions.
The eclipse maps show no evidence of a bright spot at disc rim, and no
enhanced emission along the gas stream trajectory.

While the $J$ and $K_s$ eclipse maps are dominated by emission from a
central brightness source, they display asymmetries in their outer regions
which suggest the presence of an elliptical outer disc with major axis
roughly along the direction perpendicular to the line joining both stars
(the vertical direction in Fig.~\ref{fig:maps}). This is in line with the
expected orientation of the two-armed spiral structure that may be induced
in the outer disc regions by the tides from the mass-donor star
\citep[][and references therein]{Sawadaetal86,Boffin01,Steeghs01}.
Indeed, the $H$-band light curve shows the characteristic 'bulge' in the
eclipse shape which signals the presence of a two-armed spiral structure
in an accretion disc and is reminiscent of the light curves of the
eclipsing dwarf nova IP\,Peg in outburst \citep{bhs00,Baptista2005}.
Accordingly, the corresponding eclipse map shows two clear asymmetric arcs
in the outer disc regions, located in the upper left (arm 1) and lower
right (arm 2) quadrants.  The resemblance of the $H$-band light curve and
eclipse map of V2051\,Oph with the light curves and corresponding eclipse
maps of IP\,Peg in outburst lead us to interpret the observed asymmetric
arcs as tidally-induced spiral arms in the quiescent disc of V2051\,Oph.

Two-armed spiral structure in high inclination accretion discs lead to
a double-wave orbital modulation, with maxima coinciding with the binary
phases where the spiral arms are seen face-on \citep[e.g.,][]
{Steeghs01,Baptista2005}.
The right-hand panels of Fig.~\ref{fig:maps} show the $JHK_s$ eclipse maps
of V2051\,Oph with their central regions ($R<0.2\,R_\mathrm{L1}$) masked for
a clearer view of the asymmetries in the outer disc regions. Small red
arcs indicate the azimuths of maximum emission of the double wave modulation
in each case. They are in good agreement with the azimuths where the spiral
arms in the eclipse maps are seen face-on. This provides additional support
for the interpretation that the asymmetric arcs seen in the eclipse maps
correspond to tidally-induced spiral arms. The azimuths of maximum double
wave emission rotate clockwise with increasing wavelength, suggesting that
the temperature along the spiral structure decreases towards their tails
(i.e., towards larger radii).

\begin{figure}
  \centering
  \includegraphics[width=0.48\textwidth]{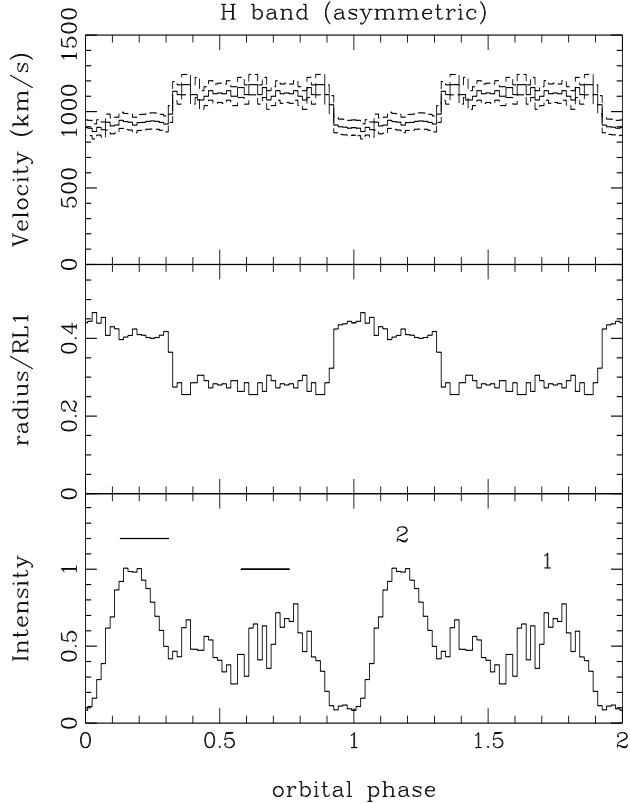}
  \caption{ Dependency with binary phase of the maximum intensity, radius
   and corresponding Keplerian velocity at maximum intensity as derived from
   the $H$-band map. Intensities are plotted in an arbitrary scale; arms 1
   and 2 are labeled and the phases of maxima of the double-wave orbital
   modulation are indicated by horizontal tick marks. Keplerian velocities
   were computed assuming $M_1= 0.78\pm 0.02\,M_\odot$ and $R_\mathrm{L1}=
   0.422 \pm 0.015 \,R_\odot$ \citep{Baptistaetal98}. Dashed lines in the
   velocity panel shows the uncertainties at the 1$\sigma$ limit. }
    \label{fig:spirals}
\end{figure}

In order to investigate the properties of the asymmetric arcs we followed
the steps of \cite{bhs00} and divided the $H$-band eclipse map into
azimuthal slices (i.e., ``slices of pizza''), computing the radius at
which the intensity is a maximum for each azimuth. A corresponding Keplerian
velocity is obtained for the radius of maximum intensity assuming
$M_1= (0.78\pm 0.06)\, M_\odot$ and $R_\mathrm{L1}= (0.422\pm 0.015)\,R_\odot$
\citep{Baptistaetal98}. This exercise allow us to recover the location
of the spiral structures both in radius and in azimuth
\citep[e.g.,][]{bhs00,Baptista2005}.
Fig.~\ref{fig:spirals} shows the azimuthal intensity distribution
$I_\mathrm{max}$, corresponding radius $R(I_\mathrm{max})$ and Keplerian
velocity $v_\mathrm{kep}[R(I_\mathrm{max})]$. Azimuths are expressed in terms
of orbital phase; these are measured from the line joining both stars and
increase clockwise. 
%
%
%
The two-armed asymmetric structures in the $H$-band eclipse map lead to a
double-humped shape in the azimuthal intensity distribution. The maxima
indicate the positions of arms 1 and 2, whereas the valleys trace the
region in between the spirals. The binary phases of maximum intensity of
the spiral arms are in good agreement with the observed phases of maximum
of the double-wave orbital modulation (indicated by horizontal tick marks
in Fig.~\ref{fig:spirals}).
Similar to IP\,Peg \citep{bhs00}, the spiral arms are located at different
distances from disc centre. The maximum intensity along the outer arm 2
occurs at a radius of $0.42 \pm 0.02 \,R_\mathrm{L1} \,(v_\mathrm{kep}=
910 \pm 50\,\, \mathrm{km\, s^{-1}})$, whereas that of the inner arm 1 is
at $0.28 \pm 0.02 \,R_\mathrm{L1} \, (v_\mathrm{kep}= 1120\pm 60\,\,
\mathrm{km\,s^{-1}})$.

As a consequence of the intrinsic azimuthal smearing effect of the
eclipse mapping method, spiral arms are smeared in the azimuthal
direction; their trace in the azimuthal intensity distribution will
always be at a constant average radius which cannot be used to estimate
the opening angle of the spirals \citep{bhs00,Harlaftis04,Baptista2005}.
Instead, \cite{Baptista2005} realized that the opening angle of the
spirals $\theta_s$ can be estimated from the orientation of the valleys
in the azimuthal intensity distribution, which rotate clockwise as the
opening angle decreases and the spirals wind up. They found that the
orbital phase of lowest intensity (the valley) in the azimuthal intensity
distribution correlates to the opening angle of the spirals through the
expression,
\begin{equation}
\theta_s(degrees) = \frac{23.25}{\phi_1 (I_\mathrm{min})} - 29.6 \, ,
\end{equation}
where $\phi_1 (I_\mathrm{min})$ corresponds to the first of the two orbital
phases of intensity valleys in the azimuthal intensity distribution
(if the spirals are diametrically opposed, $\phi_1 = \phi_2 - 0.5$).
We fitted a parabola to the intensities around the second, better defined
valley of the azimuthal intensity distribution of Fig.~\ref{fig:spirals}
and assumed diametrically opposed spiral arms
\footnote{Given that the two arcs are separated in phase by
  $\simeq 0.5$, the assumption that the spiral arms are diametrically
  opposed seems a reasonable one.}
to find $\phi_1 (I_\mathrm{min})= 0.460 \pm 0.035$, correspondind to an
opening angle of $\theta_s= 21^o \pm 4^o$.
The opening angle inferred for the spiral arms in the quiescent disc
of V2051\,Oph is in between those seen in IP\,Peg 5-6 days
\citep[$\theta_s= 25^o \pm 3^o$,][]{Baptista2005} and 8-9 days
\citep[$\theta_s= 14^o \pm 3^o$,][]{bht02} after outburst onset.

Additional evidence for the presence of tidally-induced spiral arms
in the quiescent accretion disc of V2051\,Oph comes from the
Doppler tomography studies of \cite{Papadaki08}, who observed
V2051\,Oph in quiescence two days after the end of its 1999
superoutburst, and of \cite{Rutkowski}, who framed V2051\,Oph in
quiescence several months after an outburst. The trailed spectrograms
of \citet{Papadaki08} shows double peaked emission lines, the
brightness and velocity separation of which are modulated at half the
orbital period --- a known signature of spiral arms in accretion discs
(See also Sect.~\ref{curves}). Accordingly, their Balmer and He\,I
tomograms show two asymmetric arcs extended in azimuth at opposite sides
of the accretion disc, the trace of which bends towards lower velocities
with increasing binary phase \citep[as expected for a spiral pattern
moving towards progressively larger radii with binary phase, see e.g.
Fig.\,4 of][]{ss99}. These arcs are reminiscent of the spiral structures
seen in Doppler tomograms of IP\,Peg in outburst \citep{Steeghsetal97},
but rotated in phase by $\Delta \phi \sim 0.1-0.15$.
This apparent rotation led \cite{Papadaki08} to discard the possible
interpretation of the observed structures in terms of tidally-induced
spiral arms with the argument that their orientation would be unusual.
However, spiral arms observed in Doppler tomograms of outbursting
dwarf novae appear at distinct orientations for different objects
\citep[see, e.g., Fig.\,11 of][]{Steeghs01}, and the Doppler tomography
of U\,Gem in outburst shows that the line emission from its spiral arms
does rotate significantly in azimuth as the system declines from
outburst maximum \citep{Groot01}, possibly reflecting changes in
line emissivity along its spiral pattern.
Interestingly, \cite{Pattersonetal03} performed photometric monitoring
of V2051\,Oph along the same 1999 superoutburst and found that, aside
of the superhump signal, there were significant power at
$P_\mathrm{orb}/2$, suggesting that tidally-induced spiral arms were
already present in its accretion disc during superoutburst, cohexisting
with the elliptical precessing disc.

\cite{Rutkowski} performed Doppler tomography over three
consecutive orbits of V2051\,Oph in quiescence, 150 days after the
end of the previous outburst. Their $H_\alpha$ tomogram shows a
one-armed spiral structure, while the tomograms of all other Balmer,
He\,I and O\,I lines show clear two-armed spiral structures at the
location where they are usually seen in outbursting dwarf novae
\citep[e.g.,][]{Steeghs01}, which they interpreted as tidally-induced
density waves. They estimated the disc temperatures from the Doppler
tomograms and found no enhanced temperatures at the location of the
two-armed spirals, suggesting that these structures are likely the
result of increased density instead of being caused by strong
hydrodynamical shocks.

The AAVSO database on V2051\,Oph shows that the nearest recorded outburst
occured 42 days before our observations and was apparently just a normal
outburst. Therefore, the interpretation of \cite{Papadaki08} that the
inner arm is related to the bright spot and the outer arm is reminiscent
superhump light source is neither a plausible explanation for our
observations nor for those of \cite{Rutkowski}. On the other hand,
the interpretation of the two-armed spiral structure seen in the Doppler
tomograms of \cite{Papadaki08}, \cite{Rutkowski}, and in our eclipse
maps as tidally-induced spiral arms, provides a simple and plausible
explanation for all observations.

\subsection{Spatially-resolved disc spectra}
\label{spectra}

Our broad-band $JHK_s$ eclipse maps yield spatial information about the
emerging infrared spectrum of the V2051\,Oph accretion disc. By combining
the three intensity maps we are able to recover broad-band infrared
spectra at different disc locations.

\begin{figure}
  \centering
  \includegraphics[width=0.48\textwidth]{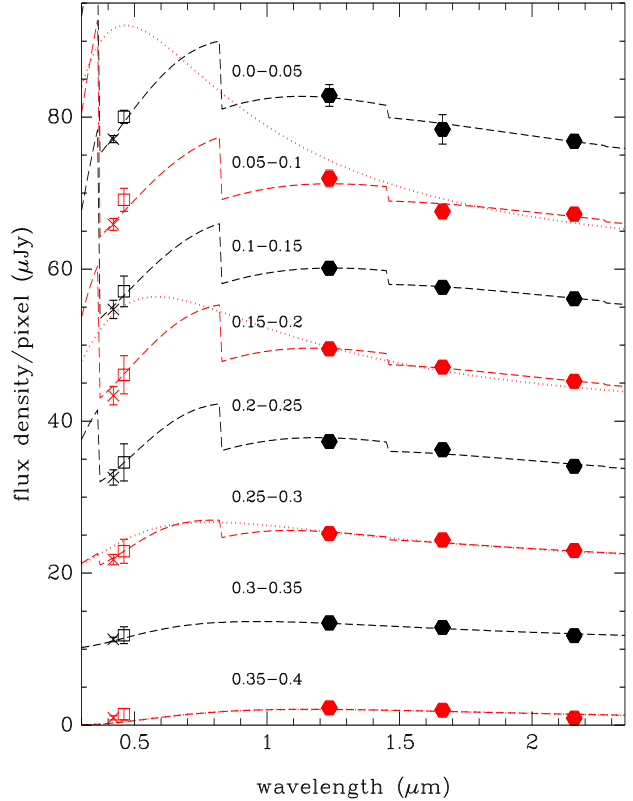}
  \caption{Spatially-resolved $JHK_s$ disc spectra computed for a set of
    concentric annular sections (radial range indicated on the left side
    in units of $R_\mathrm{L1}$). Error bars were derived via Monte Carlo
    simulations with the eclipse light curves. Except for the lowest,
    each spectrum was vertically offset by $10\,\mu Jy$ from the previous
    one for visualization purposes. Dashed lines show the best-fit LTE
    bound-free plus free-free hydrogen continuum emission spectrum in each
    case, while red dotted lines show the best-fit blackbody model in
    four cases. The best-fit models were extrapolated to the optical
    range in order to allow a direct visual comparison with the observed
    $B$-band fluxes from the 'faint' (crosses) and 'bright' (open squares)
    quiescent states of \citet{bb04}. The $B$-band fluxes were horizontally
    displaced from the central wavelength of the $B$-band for visualization
    purposes ($ 1\,\mu Jy = 10^{-29} erg\,cm^{-2} s^{-1} Hz^{-1}$).}
    \label{fig:spectra}
\end{figure}

We extracted disc spectra as a function of radius by dividing the maps
into concentric annular regions of width $0.05\,R_\mathrm{L1}$ centered at
the WD. The radial width of the regions is chosen based on the spatial
resolution and the signal-to-noise ratio of the eclipse maps. In order to
minimize the contribution of the spiral arms to the disc emission, each
spectrum is obtained by computing the median intensity of all pixels inside
the corresponding annulus; the standard deviation of the randomized eclipse
maps (see Sect.~\ref{eclipsemap}) with respect to the median intensity is
added in quadrature to the 10 per cent uncertainty from the subtraction
of the contribution from the mass donor star to the light curves, and the
result is taken as the corresponding uncertainty for each annulus. The
resulting broad-band disc spectra are shown in Fig.~\ref{fig:spectra}.
%
%
Ordinates are given in terms of flux per pixel in the eclipse map as seen
at the Earth (a quantity independent of distance).
The disc is hot and optically thin in its inner regions (with the Brackett
jump clearly in emission) and becomes cold and opaque in the outer regions,
those containing the spiral arms.

Given that there is yet no accepted, detailed accretion disc atmosphere
model which describes how temperature changes in the vertical direction,
particularly in quiescence, the following modelling of the observed disc
spectra should be looked at with some reservation. 

The flux per pixel in the eclipse maps, $f_\nu$, is related to the intrinsic
intensity of the corresponding pixel, $I_\nu$, by the expression
$f_\nu = \Theta^2 \, I_\nu $, where its solid angle is given by,
\begin{equation}
\Theta^2 =  \left( \frac{2}{N} \frac{R_\mathrm{L1}}{d} \right)^2 \, \cos i =
\left( \frac{2}{N} \frac{R_\mathrm{L1}/R_\odot}{d/Kpc} \right)^2 \cos i \,
\left[ \frac{R_\odot}{Kpc} \right]^2 \,\, ,
\end{equation}
and $N= 51\times 51$ is the total number of pixels in the eclipse map.
Adopting $R_\mathrm{L1}= (0.422 \pm 0.015)\,R_\odot$, $d= (112.3 \pm 0.9)$\,pc
and $i= 83.3^o \pm 1.4^o$ leads to $\Theta^2_D= (2.5 \pm 0.5) \times 10^{-3}\,
(R_\odot/Kpc)^2$.

Because our broad-band infrared spectra consist of only three data points,
we are bound to fitting them with spectral models of up to two free
parameters. Therefore, we fixed the distance (i.e., $\Theta^2$) and
used the synphot/IRAF
  \footnote{IRAF is distributed by the National Optical Astronomy
    Observatories, which are operated by the Association of Universities
    for Research in Astronomy, Inc., under cooperative agreement with the
    National Science Foundation.}
package to generate grids of $JHK_s$ fluxes for blackbodies (effective
temperature $T_\mathrm{ef}$ as free parameter), stellar atmosphere models
($T_\mathrm{ef}$ and $\log g$), and LTE bound-free plus free-free hydrogen
continuum emission spectra ($T_\mathrm{gas}$ and column density $N(H)$)
covering a range of input parameter values. We then computed the chi-square
between the disc spectra and the fluxes in the grid of models to find
the region of best-fit parameter combination in each case, and we further
improved the solution using a refined grid around the best-fit region.
Because of the Brackett jump in emission, blackbodies and stellar models
do not provide good fits to the spectra of the inner disc regions
($R < 0.3\,R_\mathrm{L1}$); best-fit blackbodies and hydrogen emission models
are indistinguishable in the outer, opaque disc regions ($R \geq 0.3\,
R_\mathrm{L1}$). The best fit solutions ($\chi^2_\mathrm{red} \leq 1$) are
obtained with the bound-free plus free-free hydrogen emission model in
all cases; these are shown as dashed lines in Fig.~\ref{fig:spectra}.
Uncertainties in the fitted parameters were derived from Monte Carlo
simulations with the fluxes of the spatially-resolved disc spectra,
subjecting the set of randomized spectra to the same fitting procedure.
Quoted uncertainties are the standard deviation with respect to the
average parameter value in each case.

We further tried fitting the data using blackbody spectra with $\Theta^2$
as a second free parameter, but this results in chi-square values which
are larger than and comparable to the bound-free plus free-free hydrogen
emission model fit, respectively at the inner (optically thin) and outer
(opaque) disc regions. Increasing the solid angle ($\Theta^2 > \Theta^2_D$,
implying $\cos i_\mathrm{ef} > \cos i$ and larger pixel effective surface)
does not improve the quality of the fits. Therefore, we found no evidence
of vertically extended emission anywhere in the disc. 
%
\begin{figure}
  \centering
  \includegraphics[width=0.37\textwidth,angle=-90]{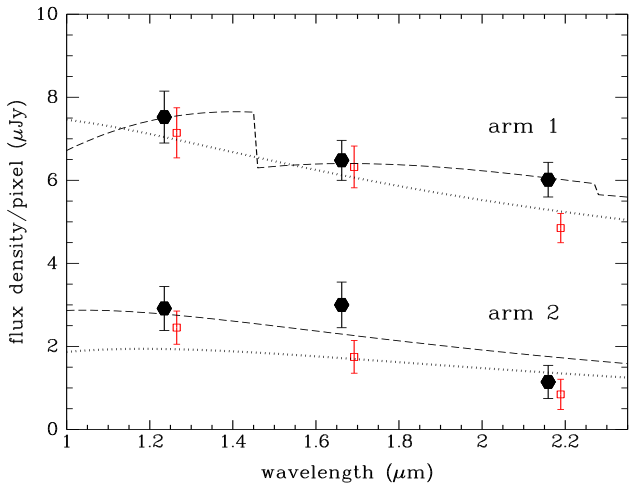}
  \caption{Spatially-resolved $JHK_s$ spectra of the spiral arms (filled
    circles) and of the disc at the same radial range (open squares). The
    notation is similar to that of Fig.~\ref{fig:spectra}. The best-fit
    corresponding LTE bound-free plus free-free hydrogen continuum emission
    models are shown as dashed (spiral arms) and dotted (disc) lines. }
    \label{fig:arms}
\end{figure}

The inner disc atmosphere is made of hot ($T_\mathrm{gas} \sim 10-12
\times 10^3\,K$) and tenuous gas (with corresponding surface densities
$m_H\,N(H) \sim 10^{-3}-10^{-2}\,g\,cm^{-2}$, where $m_H$ is the
hydrogen mass). The inferred surface densities are several orders of
magnitude smaller than those predicted by the DIM
\citep[$\sim 10^1 - 10^2\,g\,cm^{-2}$, see e.g.][]{Lasota2001} and those
expected for viscous, steady-state discs at low mass accretion rates
\citep[$\sim 0.5-1 \,g\,cm^{-2}$ for $\alpha=1$ and \.{M}$= 10^{-11}\,
M_\odot\,yr^{-1}$,][]{acpower}, and can only be matched by unrealistically
large disc viscosities ($\alpha > 10^4$) or unrealistically low mass
accretion rates of $10^{-16}-10^{-14}\,M_\odot\,yr^{-1}$.
Our results suggest that either an extended hot and tenuous chromosphere
veils the emission from an underlying, colder accretion disc up to
$R \sim 0.3\,R_\mathrm{L1}$, or that the inner disc itself has evaporated
into a hot and tenuous corona --- although such quiescent accretion disc
corona are expected to be much hotter than the $\sim 10^4\,K$ inferred
from our study \citep[e.g.,][]{Meyer00,hellier,Barbera17}.
These results are in good agreement with and underscore the previous
findings of \cite{Vrielmann02} and \cite{Saito06} --- which suggested
that the quiescent accretion disc of V2051\,Oph is a sandwich of a cool,
optically thick underlying disc with hot chromospheric layers on both
sides ---, and of \cite{Berriman86}, which found that most of the
V2051\,Oph disc would consist of hot and optically thin gas for binary
distances above 100\,pc (as confirmed by Paper\,I and the results of
Sect.~\ref{distance}).

Fig.\ref{fig:spectra} shows the $JHK_s$ best-fit models extrapolated to
the optical range in order to allow a direct visual comparison with the
observed $B$-band fluxes derived by \citet{bb04} from their 'faint'
(crosses) and 'bright' (open squares) V2051\,Oph quiescent states.
We remark that these optical observations were not included in the
fitting process, as they correspond to a different epoch than that of
our data. Nevertheless, the $JHK_s$ best-fit hydrogen continuum emission
models are in very good agreement with the $B$-band observations of
\citet{bb04}, suggesting that the inner disc regions were also
optically thin and with a similar radial temperature distribution
during their observations.
We note that such good agreement weakens the idea of an underlying
cooler and opaque inner disc, as neglecting such a component in the
spectral fit would lead to predicted $B$-band model fluxes which would
systematically deviate from the observations. Our $JHK_s$ best-fit
hydrogen continuum emission models also predict strong Paschen and
Balmer jumps in emission in the inner and intermediate disc regions,
in line with the findings of \citet{Saito06} -- although it is worth
noting that their observations correspond to the $B\simeq 16.2$~mag
unusual, low brightness state reported by \citet{Baptistaetal98}.

Red dotted lines in Fig.\ref{fig:spectra} show the best-fit blackbody
model in four cases. They largely overestimate the $B$-band observations
of \citet{bb04} in the optically thin disc regions and only provide a
reasonable match to their data in the outer, optically thick disc
regions where blackbodies are indistinguishable from the best-fit
hydrogen continuum emission models.

We investigated the emission from the spiral arms by computing average
$JHK_s$ fluxes at annular regions of radial width $0.1\,R_\mathrm{L1}$
and azimuthal extent $60^o$ around the position of each arm, and we
also extracted comparison disc spectra at the same radial range but
complementary azimuths (median fluxes, as before). The increased radial
width compensates for the limited azimuthal extent and helps to preserve
the signal-to-noise ratio of the spiral arms spectra. The resulting
spectra are shown in Fig.~\ref{fig:arms} as solid circles (spiral arms)
and open squares (disc at same radial range). The fluxes of the spiral
arms are systematically larger than those of the disc at same radial
range, although the differences are within the 2-$\sigma$ level.
Modelling of these spectra with LTE bound-free plus free-free hydrogen
continuum emission models indicate the spiral arms are hotter and have
lower column densities than the disc at same radial range, but that the
differences in $T_\mathrm{gas}$ and $N(H)$ are also within the 2-$\sigma$
level. Arm~1 is well described by a model with $T_\mathrm{gas}= 7300 \pm
1100 \,K$ and $\log N(H)= 24.1 \pm 2.1$, while arm~2 is reasonably well
matched by a model with $T_\mathrm{gas}= 4900 \pm 250 \,K$ and
$\log N(H)= 29.8 \pm 0.6$.  We note that for $\log N(H) \geq 25$ the
column density becomes meaningless and only indicates that the gas is
optically thick. The best-fit models are shown in Fig.~\ref{fig:arms}
as dashed (spiral arms) and dotted (disc) lines. Blackbody fits with
$T_\mathrm{ef}$ and $\Theta^2$ as free parameters show no evidence for
perceptible vertical extension of the spiral arms. The relatively low
temperatures inferred for the spiral arms in the V2051\,Oph quiescent
accretion disc suggest they are not the result of strong hydrodynamical
shocks -- in line with the results of \cite{Rutkowski}.

\subsection{The disc radial temperature distribution}
\label{radtemp}

\begin{figure}
  \centering
  \includegraphics[width=0.37\textwidth,angle=-90]{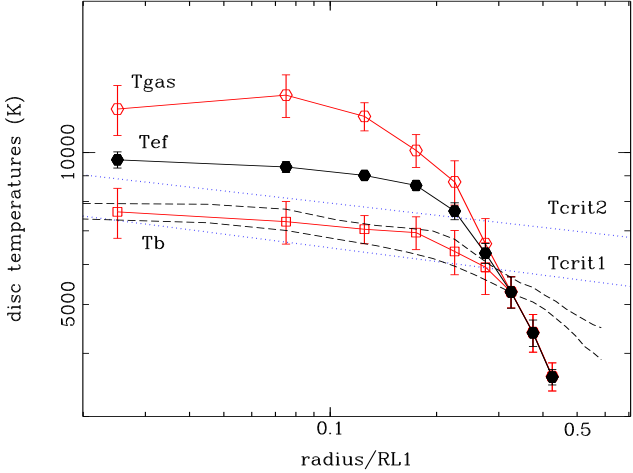}
  \caption{Radial temperature distributions derived from the
    spatially-resolved $JHK_s$ disc spectra. Gas temperatures are shown
    as open circles, corresponding effective temperatures are shown
    as filled circles, while predicted B-band blackbody brightness
    temperatures from the same models are depicted as open squares.
    The dashed lines show the B-band blackbody temperatures inferred by
    \citet{bb04}. Dotted lines show the two critical effective
    temperatures which bind the thermal-viscous disc instability. }
    \label{fig:radtemp}
\end{figure}

Fig.~\ref{fig:radtemp} shows the radial gas temperature distribution
derived from the accretion disc best-fit hydrogen continuum
emission models of Fig.~\ref{fig:spectra} (open circles).
%
%
In order to allow the comparison of the results with the $T_\mathrm{crit1}$
and $T_\mathrm{crit2}$ critical temperatures which bind the thermal-viscous
disc instability, we computed the effective temperature of the best-fit
model at each radius from the corresponding surface flux integrated over
frequency,
\begin{equation}
\sigma T_\mathrm{ef}^4 = \pi \int_0^\infty I_\nu \, d\nu \,\, .
\end{equation}
These are shown in Fig.\ref{fig:radtemp} as filled circles. We further
calculated the B-band fluxes predicted by the best-fit model at each
radius and computed the corresponding B-band blackbody brightness
temperatures. These are shown in Fig.~\ref{fig:radtemp} as open squares.
The optically thin inner disc regions have a relatively flat temperature
distribution with $T_\mathrm{ef} \sim 9-10 \times 10^3\,K$; temperatures
drop fast with increasing radius in the outer, opaque disc regions.
The inferred brightness temperatures are systematically lower than the
corresponding effective temperatures by as much as 20 per cent in the
optically thin inner disc regions, but provide accurate estimates
of $T_\mathrm{ef}$ in the outer, opaque disc regions.
The brightness temperatures predicted from our models are in very good
agreement with those inferred by \cite{bb04} from their quiescent
B-band observations of V2051\,Oph (shown as dashed lines in
Fig.~\ref{fig:radtemp}, and scaled to the 112\,pc distance estimate
of Sect.~\ref{distance}), in line with results discussed in
Sect.~\ref{spectra}. Near the outer disc radius, $R_d \sim
0.4\,R_\mathrm{L1}$, the $B$-band brightness temperatures predicted
by the $JHK_s$ best-fit model decline faster with radius than those
inferred from the observations of \citet{bb04} reflecting the fact
that the accretion disc extended to a larger radius at the epoch of
their observations.

The comparison of brightness ($T_\mathrm{b}$) and effective
($T_\mathrm{ef}$) temperatures indicate that the former may underestimate
the latter by a significant amount in optically thin disc regions. Thus,
we note that if the accretion disc of V2051\,Oph remains optically thin
during outbursts, the inferred B-band disc brightness temperatures at
outburst maximum of \cite{Baptistaetal07} may be similarly underestimated,
which could raise the inferred $T_\mathrm{b}\sim 8000\,K$ up to $T_\mathrm{ef}
\sim 10000\,K > T_\mathrm{crit2}$, thereby removing the discrepancy between
the V2051\,Oph inferred disc temperatures and the predictions of the DIM.

On the other hand, if there is no underlying cool disc, the inferred
$T_\mathrm{ef}$ values well above $T_\mathrm{crit2}$ bring the inner
quiescent accretion disc of V2051\,Oph safely to the hot stable branch
and excludes the possibility of thermal-viscous driven outbursts at these
regions. This may imply in an additional discrepancy between V2051\,Oph
and DIM predictions, as this dwarf nova shows inside-out outbursts which
seem to start near the circularization radius \citep{Baptistaetal07},
well within the hot, optically thin inner disc regions.

\section{Spiral arms in a viscous, quiescent accretion disc?}

Numerical simulations by \cite{Sawadaetal86} show that accretion discs
in compact binaries are prone to the appearance of a two-armed spiral
density wave at their outer regions, induced by the tides from the
mass-donor star. Observational confirmation of this effect first came
from Doppler tomography of emission lines of the dwarf nova IP\,Peg in
outburst by \cite{Steeghsetal97}, later underscored by detection of
similar spiral structures in other outbursting dwarf novae \citep[]
[and references therein]{MarshSchwope16}.
The observations indicate these are wide open spiral arms with
$\theta_s \simeq 15^o-30^o$ \citep[e.g.,][]{Baptista2005}.

In an inviscid accretion disc, the opening angle of the spiral arms
is of the order of $\tan\theta_s = c_s/v_\phi$, where $c_s\simeq 10\,
(T/10^4 K)^{1/2}\, km\,s^{-1}$ is the sound speed, $v_\phi \simeq 10^3\,
km\,s^{-1}$ is the disc Keplerian velocity, and $T$ is the mid-plane
disc temperature \citep[e.g.,][]{Boffin01}. This led \cite{Savonije94}
to predict that tidally-induced spiral arms would be tightly wound
and could not be detected in the relatively cool ($T \leq 10^5\,K$) and
geometrically thin ($c_s \ll v_\phi$) accretion discs of dwarf novae.
Indeed, numerical simulations of a relatively low-viscosity $\alpha=
0.05-0.1$ disc by \cite{Godon98} confirmed the presence of
tidally-induced spiral arms but at much smaller opening angles than
observed, and that unrealistically high temperatures ($\simeq 10^6\,K$)
were required in order to reproduce the wide open spiral arms seen in
the IP\,Peg outbursting accretion disc.
Additional simulations by \cite{Stehle99}
confirmed that low-viscosity discs ($\alpha=0.01$) lead to tightly
wound spiral arms, and revealed that wide open spiral arms arise when
the disc viscosity is increased ($\alpha=0.3$).  The comparison of
these simulations tells us that it is possible to match the wide open
spiral arms obtained by the low-viscosity and extremely hot disc model
of \cite{Godon98} at temperatures an order of magnitude lower just by
increasing the disc viscosity by a factor of 3-6 \citep{Stehle99}.
Therefore, while increasing the disc temperatures increases the
opening angle of the resulting spiral arms \citep[e.g.,][and
references therein]{Makita}, increasing the disc viscosity seems to
have a similar (and seemingly stronger) effect.

The point we make here is that wide open spiral arms may be more a
signature of a high viscosity disc than of a hot disc, and that a large
disc viscosity might lead to detectable, wide open tidally-induced
spiral density waves even when the disc is relatively cool and small.
Thus, the detection of spiral arms in V2051\,Oph during
quiescence suggests that its quiescent accretion disc has high viscosity.
This is in line with the results of \cite{bb04} --- which inferred
a high viscosity $\alpha\simeq 0.1-0.2$ for the quiescent accretion
disc of V2051\,Oph --- and with those of \cite{Baptistaetal07} -- which
suggest that the outbursts of V2051\,Oph are the result of bursts
of enhanced mass transfer onto a high-viscosity accretion disc.

\section{summary}

Infrared $JHK_s$ quiescent light curves of V2051\,Oph after removal of the
ellipsoidal contribution from its mass donor star show an out-of-eclipse
double-wave modulation the amplitude of which increases with wavelength.
Eclipse maps derived from these light curves reveal two asymmetric arcs
extended in azimuth at the intermediate/outer disc regions, which are
reminiscent of those seen in eclipse maps of the dwarf nova IP\,Peg in
outburst and are similarly interpreted as tidally-induced spiral density
waves in the V2051\,Oph accretion disc. In agreement with this
interpretation, the maxima of the double-wave modulation coincide with
the binary phases where the spiral arms are seen face-on.
The spiral arms are located at different distances from disc centre;
the inner arm is at an average radius of $0.28 \pm 0.02 \,R_\mathrm{L1}$
(at a corresponding Keplerian velocity of $v_\mathrm{kep}= 1120\pm 60\,\,
\mathrm{km\,s^{-1}}$), while the outer arms is at $0.42 \pm 0.02
\,R_\mathrm{L1} \,(v_\mathrm{kep}= 910 \pm 50\,\,\mathrm{km\, s^{-1}})$.
We estimate the opening angle of these spirals to be $\theta_s= 21^o
\pm 4^o$. The detection of wide open spiral arms in V2051\,Oph during
quiescence suggests that its quiescent accretion disc has high viscosity.

Broad-band $JHK_s$ disc spectra were extracted for concentric annular disc
regions. The inner disc atmosphere ($R \leq 0.3\,R_\mathrm{L1}$) is made of
hot and tenuous gas ($T_\mathrm{gas}\sim 10-12 \times 10^3\,K$ and surface
densities $\sim 10^{-3}-10^{-2}\,g\,cm^{-2}$), suggesting that either an
extended hot and tenuous chromosphere veils the emission from an
underlying, colder accretion disc or that the inner disc itself has
evaporated into a hot and tenuous corona. The disc gas becomes cold and
opaque in its outer regions ($R > 0.3\,R_\mathrm{L1}$). The spiral arms
are hotter and have lower column densities than the disc at same radial
range, with differences within the 2-$\sigma$ confidence level. The
$JHK_s$ best-fit hydrogen continuum emission  models are in very good
agreement with the $B$-band observations of \citet{bb04}, suggesting
that the inner disc regions were also optically thin and with a
similar radial temperature distribution during their observations.

$B$-band blackbody brightness temperatures computed from our $JHK_s$
best-fit hydrogen continuum emission models are systematically
lower than the corresponding effective temperatures by as much
as 20 per cent in the optically thin inner disc regions, but
provide accurate estimates of $T_\mathrm{ef}$ in the outer, opaque disc
regions. If the accretion disc of V2051\,Oph is optically thin at
outbursts, inferred $B$-band disc brightness temperatures at outburst
may be similarly underestimated and the disc effective
temperatures could be higher than $T_\mathrm{crit2}$, thereby removing
the discrepancy between the V2051\,Oph inferred disc temperatures
and predictions of the DIM \citep{Baptistaetal07}. On the other hand,
in case there is no underlying cool disc, the inferred $T_\mathrm{ef}$
values well above $T_\mathrm{crit2}$ bring the inner quiescent accretion
disc of V2051\,Oph safely to the hot stable branch and excludes the
possibility of thermal-viscous driven outbursts at this binary.

\section*{Acknowledgements}

We thank an anonymous referee for interesting discussions which helped
to improve the presentation of the results.  E.\ W.\ acknowledges
financial support from CAPES (Brazil) and CNPq (Brazil). In this
research we have used, and acknowledge with thanks, data from the
AAVSO International Database that are based on observation collected
by variable star observers worldwide.


\bibliographystyle{mnras}
\bibliography{bibliography} 


\bsp	
\label{lastpage}
\end{document}